# Effects of Lower Symmetry and Dimensionality on Raman Spectra in 2D WSe$_2$

Xin Luo[1,†], Yanyuan Zhao[2,†], Jun Zhang[2], Minglin Toh[4], Christian Kloc[4], Qihua Xiong[2,3,*], Su Ying Quek[1,+,*]

[1]Institute of High Performance Computing, 1 Fusionopolis Way, #16-16 Connexis, Singapore 138632

[2]Division of Physics and Applied Physics, School of Physical and Mathematical Sciences, Nanyang Technological University, 21 Nanyang Link, Singapore 637371

[3]Division of Microelectronics, School of Electrical and Electronic Engineering, Nanyang Technological University, Singapore, 639798

[4]School of Materials Science and Engineering, Nanyang Technological University, Nanyang Avenue, Singapore 639798

*: To whom correspondence should be addressed. Email address: phyqsy@nus.edu.sg (theory); qihua@ntu.edu.sg (experiment)
+: Present address: Department of Physics, Graphene Research Centre and Centre for Computational Science and Engineering, National University of Singapore, 2 Science Drive 3, Singapore 117551
†: These authors contributed equally to this work.

ABSTRACT:

We report the observation and interpretation of new Raman peaks in few-layer tungsten diselenide (WSe$_2$), induced by the reduction of symmetry going from 3D to 2D. In general, Raman frequencies in 2D materials follow quite closely the frequencies of corresponding eigenmodes in the bulk. However, while the modes that are Raman active in the bulk are also Raman active in the thin films, the reverse is not always true due to the reduced symmetry in thin films. Here, we predict from group theory and density functional calculations that two intra-layer vibrational modes which are Raman inactive in bulk WSe$_2$ in our experimental configuration become Raman active in thin film WSe$_2$, due to reduced symmetry in thin films. This phenomenon explains the Raman peaks we observe experimentally at ~310 cm$^{-1}$ and 176 cm$^{-1}$ in thin film WSe$_2$. Interestingly, the bulk $B_{2g}^1$ mode at ~310 cm$^{-1}$ that is Raman inactive can in fact be detected in Raman measurements under specific wavelengths of irradiation, suggesting that in this case, crystal symmetry selection rules may be broken due to resonant scattering. Both theory and experiment indicate that the $E_{2g}^1$ and $B_{2g}^1$ modes blue-shift with decreasing thickness, which we attribute to surface effects. Our results shed light on a general understanding of the Raman/IR activities of the phonon modes in layered transition metal dichalcogenide materials and their evolution behavior from 3D to 2D.

## I. INTRODUCTION

Transition metal dichalcogenide (TMD) thin films have attracted tremendous attention recently because of their unique thickness-dependent properties and potential for novel applications.[1-4] Bulk layered TMD materials, such as MoS$_2$ and WSe$_2$, consist of slabs of trilayers (TL; three atomic layers) with ionic-covalent metal-chalcogenide bonds within each TL. Inter-TL interactions are dominated by weak van der Waals (vdW) interactions.[5] While bulk TMDs have been studied for a long time, the recent focus on TMD thin films has uncovered numerous intriguing phenomena that highlight the importance of symmetry breaking and quantum confinement when moving from 3D to 2D. For example, recent experiments have found that there is an indirect-to-direct bandgap transition when the thickness is reduced to single TL,[6-8] an effect which arises from an increase in the indirect band gap due to quantum confinement and the absence of inter-layer interactions in the single TL.[7] Furthermore, despite the weak strength of the vdW interactions, TMD thin films have subtle surface effects that have been shown to give rise to anomalous frequency trends in the Raman-active in-plane $E_{2g}^1$ mode.[9] Thin film TMD materials also exhibit very different symmetry properties from their bulk counterparts. For instance, exotic low-frequency Raman spectra have been reported in few-TL TMDs as a result of the reduced symmetry in 2D,[10] and the symmetry variations can be detected by the optical second harmonic generation method.[11,12] In single TL TMDs, due to the absence of inversion symmetry, spin-orbit coupling results in a coupling between the spin and valley degrees of freedom leading to potential valleytronic applications.[13-18]

In this work, we focus on how symmetry and low dimensionality influence the Raman phonon spectra of the prototypical layered TMD material, WSe$_2$. Raman spectroscopy is a non-destructive, powerful technique for characterizing the zone-center phonon properties in the bulk and at the nanoscale. The Raman frequencies in layered TMDs are very sensitive to the thickness of the thin films, making Raman spectroscopy a useful tool to determine the film thickness of these materials.[10] On the other hand, Raman activity is governed by selection rules determined by the symmetry properties of the sample, which are different for bulk and thin film materials. While modes that are Raman active in the bulk are also Raman active in the thin films, the reverse is not always true due to the reduced symmetry in thin films.

The Raman spectra of WSe$_2$ thin films are particularly interesting for this study as previous experiments have uncovered several peaks that cannot be explained by the corresponding bulk spectra. Specifically, Raman spectroscopy at 80 K has shown that the high frequency spectrum for bulk WSe$_2$ has peaks at 253, 250 and 178 cm$^{-1}$, corresponding to the Raman-active $A_{1g}$, $E_{2g}^1$, and $E_{1g}$ modes respectively.[19] This result was confirmed by a recent high-resolution Raman spectroscopy performed at room temperature, which found that the frequencies of $A_{1g}$ and $E_{2g}^1$ modes are 250.8 and 248.0 cm$^{-1}$, respectively.[20] On the other hand, for few-TL WSe$_2$, experiments observed a peak near 250 cm$^{-1}$ and a broad shoulder near 260 cm$^{-1}$ (temporarily assigned as $E_{2g}^1$ and $A_{1g}$ modes respectively) as well as an unexplained small Raman peak near 310 cm$^{-1}$.[20,21] However, under such assignments, the frequency trends for the $A_{1g}$ mode would be opposite to that expected from other similar TMD materials.[9,22]

Here we present both theoretical and experimental Raman scattering studies on the lattice vibrations in few-TL and bulk WSe$_2$. Based on our DFT calculations and group theory analysis, the frequency difference between the $A_{1g}$ and $E_{2g}^1$ modes is smaller in thin films than in the bulk, and the two modes cannot be distinguished by Raman measurements on thin films due to the weak intensity of the $E_{2g}^1$ mode[23] (strictly speaking, the notations of those modes are not $A_{1g}$ and $E_{2g}^1$ in thin films; the correct notations are given in Table 1). Our results also show that the reduction of symmetry in thin films will greatly change the Raman activities of the vibration modes, resulting in the appearance of additional Raman peaks in $N$TL ($N \geq 2$) WSe$_2$.

The surface effect is found to play an important role in the thickness dependent frequency evolution trends of both in-plane $E_{2g}^1$ and out-of-plane $B_{2g}^1$ modes. Based on this analysis, not only can we explain the previously published Raman spectra,[10] but also, we confirm the theoretical predictions using different polarization configurations in our Raman experiments, as well as uncover new peaks that have been overlooked in previous experiments.

## II. METHODOLOGY

The *ab initio* calculations are performed using density functional theory (DFT) as implemented in the plane-wave code QUANTUM-ESPRESSO.[24] DFT calculations within the local density approximation (LDA) are known to give good descriptions of the atomic structure and phonon properties in layered materials, even though LDA does not take into account the van der Waals interactions.[10,25-28] Therefore, we compute the phonon spectra and Raman intensities using LDA, within density functional perturbation theory,[29] using projector-augmented wave potentials for the frequency calculations and norm-conserving potentials for the Raman intensities. To get converged results, we use an energy cutoff of 65 Ry, and Monkhorst-Pack k-point meshes of 17×17×5 and 17×17×1 for bulk and thin film systems, respectively. A vacuum thickness of 16 Å is used to separate neighboring slabs in the thin film calculations. The structures are fully relaxed until all forces are less than 0.003 eV/Å. Based on the LDA calculated frequencies of bulk and 1TL WSe$_2$, a force constants model with interactions up to the second nearest neighbor is constructed to understand the frequency trends.

Bulk single crystals of WSe$_2$ were grown by the vapor phase transport method. The samples of few-TL WSe$_2$ are exfoliated from bulk WSe$_2$ crystals onto the 100 nm SiO$_2$/Si substrates using the conventional mechanical exfoliation technique.[3] Raman scattering spectroscopy measurements are carried out at room temperature using a micro-Raman spectrometer (Horiba JY-T64000) in a backscattering configuration. A solid-state laser (λ=532 nm) and an Ar ion laser (λ=488 nm) have been used to excite the samples. The backscattered signal was collected through a 100X objective and dispersed by a 1800 g/mm grating. To evaluate the impact from substrate, we have also performed Raman measurements on suspended WSe$_2$ thin films. These films were exfoliated and transferred to SiO$_2$/Si substrates with premade holes (the hole size is ~2μm) using a well-established PMMA transfer method.[30]

### III. RESULTS AND DISCUSSION

Bulk WSe$_2$ crystallizes in the 2H structure with a primitive unit cell consisting of two trilayers (TL, one trilayer contains three atomic planes) with the atomic layers arranged in /AbA BaB/ stacking as shown in Figure 1. Within each TL, W and Se share ionic-covalent bonds, while adjacent TLs interact via weak vdW interactions. The LDA optimized in-plane ($a$) and out-of-plane ($c$) lattice constants of bulk are 3.256 Å and 12.839 Å, respectively, within 0.9% deviation from the experimental values ($a$=3.282 Å and $c$=12.96 Å).[31] The in-plane lattice constant of optimized thin film ($a$=3.253Å in 1TL) is slightly smaller than that of the bulk, and will approach the bulk value as the thickness increases. Interestingly, from our calculations on 2-7TL WSe$_2$, we find that the surface W-Se valence bond length is shorter than the interior one by 0.1%, a similar surface contraction has also been observed in MoS$_2$ thin films based on low energy electron diffraction studies.[32]

The symmetry of 2H WSe$_2$ can be described by the $D_{6h}^4$ (P6$_3$/mmc) space group, and the irreducible representation of phonon modes at the centre of the Brillouin zone is $\Gamma_{bulk} = A_{1g} + 2A_{2u} + B_{1u} + 2B_{2g} + E_{1g} + 2E_{1u} + E_{2u} + 2E_{2g}$, among which there are four Raman-active modes ($A_{1g}$, $E_{1g}$, $E_{2g}^1$, $E_{2g}^2$), four infrared-active modes ($E_{1u}^1$, $E_{1u}^2$, $A_{2u}^1$, $A_{2u}^2$) and four optically inactive modes ($E_{2u}$, $B_{2g}^2$, $B_{1u}$, $B_{2g}^1$). The two-fold degenerate $E$ modes represent in-plane vibrations and the non-degenerate $A$ modes represent out-of-plane vibrations. When the system goes from 3D to 2D, the translation symmetry along the $z$-axis is absent. As a result, films with an odd number of TLs (odd-TLs) belong to the space group $D_{3h}^1$ ($P\bar{6}m2$) without inversion symmetry, while films with an even number of TLs (even-TLs) belong to the space group $D_{3d}^3$ ($P\bar{3}m1$) with inversion symmetry.[10] According to group theory, their corresponding irreducible representations of the zone center phonons can be written as:

$$\Gamma_{odd-TLs} = \frac{(3N-1)}{2}\left(A_1' + A_2'' + E' + E''\right) + A_2'' + E', \ N = 1, 3, 5\ldots,$$

$$\Gamma_{even-TLs} = \frac{3N}{2}\left(A_{1g} + A_{2u} + E_g + E_u\right), \ N = 2, 4, 6\ldots$$

where $A_1'$, $E''$, $A_{1g}$ and $E_g$ are Raman active modes, $A_2''$, $A_{2u}$ and $E_u$ are infrared active modes, and the $E'$ is both Raman and infrared active.

Table 1 shows the LDA calculated $\Gamma$ point phonon frequencies in bulk and 1-4TL WSe$_2$, together with the correct notations from group theory analysis. The non-resonant Raman intensity is calculated within the Placzek approximation,[33] and it is proportional to $\left|e_i \cdot \widetilde{R} \cdot e_s\right|^2$, where $e_i$ and $e_s$ are polarization vectors of the incident and scattered light, respectively, and $\widetilde{R}_{ij}$ is the second rank polarizability tensor. To compare with the experimental Raman spectra, we compute the Raman intensities $I_{xx}$ (in parentheses of Table 1) where the polarizations of both incident and scattering light are parallel. The Raman intensity $I_{xy}$ (the polarizations of incident light and the scattered light are perpendicular) has the same value as $I_{xx}$ for all the in-plane vibrational modes, and gives zero values for all the out-of-plane vibrational modes, a finding that is consistent with the group theory analysis. We further note that the frequency differences between experimental values and LDA calculated results are within 2 cm$^{-1}$ for the $A_{1g}$, $E_{2g}^1$, and $E_{1g}$ Raman modes in bulk WSe$_2$.

For few-TL WSe$_2$, we found that each mode in the single TL evolves into $N$ modes in $N$ TLs with similar frequencies. The ultra-low frequency evolution trends have been studied in our previous work,[10] and we focus on the high frequency range in this paper. As has been observed in other TMD thin films,[7,20,22,34-36] the Raman active modes with large intensity show some thickness-dependent trends, for example, the $E_{2g}^1$ modes blue-shift and $A_{1g}$ modes red-shift with decreasing thickness. As a result, the computed frequency differences of the two modes are 1.1 and 2.6 cm$^{-1}$ in 1TL ($E'$ and $A_1'$ in Table 1) and 2TL ($E_g$ and $A_{1g}$ in Table 1), respectively. Since the calculated frequency difference of $E_{2g}^1$ and $A_{1g}$ in bulk is 4.3 cm$^{-1}$ compared with 2.8 cm$^{-1}$ in experiment, the experimental frequency difference of corresponding modes in few-TLs will be smaller than in bulk. This small frequency difference in the few-TLs implies that the two modes would be hard to distinguish in the Raman spectra, explaining the presence of only one peak near 250 cm$^{-1}$ in few-TL WSe$_2$ under 532 nm excitation, as shown in Figure 2. In fact, under certain wavelength excitation, the $E_{2g}^1$ and $A_{1g}$ modes are still distinguishable - Figure S1 shows the weak signal of both peaks under 633 nm excitation.[37]

On the other hand, our experiments also find a peak at 260 cm$^{-1}$, which has different assignments in the literature. Some literature tentatively attributed it to the $A_{1g}$ mode,[10,11,21,38] while others assigned it to second order Raman processes.[20,23] As described above, the $A_{1g}$ frequency should be much closer to the $E_{2g}^1$ frequency, and we believe that this peak at 260 cm$^{-1}$ should not be the $A_{1g}$ mode. Further evidence comes from our Raman measurements using different polarizations of the incident and scattered light. Our experimental Raman scattering geometries are represented by the Porto notations[39] $\bar{z}(xx)z$ and $\bar{z}(xy)z$, where $\bar{z}$ and $z$ indicate wave vectors of incident and scattered light. The two alphabets in parentheses represent the polarizations of incident and scattered light, respectively. The peak at 260 cm$^{-1}$ is observed under both $\bar{z}(xx)z$ and $\bar{z}(xy)z$ polarized configurations (Fig. 2a and 2c), in contrast to that expected for the out-of-plane $A_{1g}$ mode, which according to group theory and our calculations, has zero Raman intensity under the $\bar{z}(xy)z$ polarized configuration. Specifically, the Raman tensor for the $A_{1g}$ mode relevant for bulk and even-TLs and the corresponding $A_1'$ mode in odd-TLs is

$$A_{1g} = A_1' = \begin{pmatrix} a & 0 & 0 \\ 0 & a & 0 \\ 0 & 0 & b \end{pmatrix},$$

thus giving zero intensity under the $\bar{z}(xy)z$ polarization configuration. Therefore, the large intensity of the unknown mode near 260 cm$^{-1}$ under the $\bar{z}(xy)z$ configuration suggests that it is not an out-of-plane vibrational mode. It has been discussed that uniaxial strains can split some degenerate in-plane vibration modes and give rise to new peaks.[38] To exclude the effect of strain in our experiments, we perform the Raman scattering on suspended samples and still observe the peak near 260 cm$^{-1}$, as shown in Figure 2d. This peak may come from second order Raman processes,[20,23,34] which is out of the scope of the present paper.

Table 1 also suggests the possibility of new peaks in $N$TL ($N \geq 2$) WSe$_2$ due to the lower symmetry in thin films. For instance, the bulk optically-inactive $B_{2g}^1$ mode near 310 cm$^{-1}$ will evolve into the Raman-active $A_{1g}$ mode in even-TLs (311.3 and 310.1 cm$^{-1}$ in 2TL and 4TL respectively) and the Raman-active $A_1'$ mode in odd-TLs for $N > 1$ (311.0 cm$^{-1}$ in 3TL). The corresponding mode at 310 cm$^{-1}$ in 1TL belongs to the $A_2''$ group and is not Raman-active. Modes in 1TL evolve into $N$ modes with similar frequencies in $N$ TLs and in this case, the infrared-active 1TL mode at 312.6 cm$^{-1}$ evolves into two infrared-active $A_2''$ modes and one Raman-active $A_1'$ mode in 3TL close to 310 cm$^{-1}$. In particular, this Raman-active $A_1'$ mode differs from the corresponding $A_2''$ mode in 1TL in that the mode is symmetric about the middle atomic plane, that is invariant. Similarly, in 5TL, there are 3 infrared-active and 2 Raman-active modes close to 310 cm$^{-1}$ (see Fig. S2).[37] Furthermore, our calculations predict that this mode is an out-of-plane vibration mode with zero Raman intensity in the $\bar{z}(xy)z$ polarization configuration but finite Raman intensity in the $\bar{z}(xx)z$ polarization configuration. These predictions are confirmed by our Raman measurements, in which a peak at 310 cm$^{-1}$ is observed in $N$TL ($N \geq 2$) under the $\bar{z}(xx)z$ polarization configuration only (Fig. 2a and 2c). As predicted, this peak is missing from the 1TL case. However, we find, surprisingly, that the peak exists in the bulk spectra as well, as can be seen from the zoom-in spectra in Fig. 2b. We believe that this peak may result from a breaking of Raman selection rules due to resonant Raman processes,[40] as no such peak is present when we change the incident laser to 488 nm (Fig. 3c). Fig. 2b clearly shows that the peak frequency exhibits a slight blue shift with decreasing thickness, which can be well explained by our calculations and will be discussed in detail later.

Besides the $B_{2g}^1$ mode, our calculations also suggest that the $E_{1g}$ mode (176.3 cm$^{-1}$), which is Raman active in bulk but has zero intensity under the backscattered $\bar{z}(xx)z$ and $\bar{z}(xy)z$ polarization configurations, may have finite intensity and could be observed in $N$TL ($N \geq 2$) WSe$_2$ under the same $\bar{z}(xx)z$ polarization configuration due to the reduction of symmetry. Note that the bulk $E_{1g}$ mode originates from $E''$ in 1TL, and will evolve into $E_g$ in 2TL and $E'$ mode in 3TL. Their corresponding Raman tensors are

$$E_{1g} = E'' = \begin{pmatrix} 0 & 0 & a \\ 0 & 0 & b \\ a & b & 0 \end{pmatrix}, E_g = \begin{pmatrix} a & c & d \\ c & -a & f \\ d & f & 0 \end{pmatrix}, E' = \begin{pmatrix} a & c & 0 \\ c & -a & 0 \\ 0 & 0 & 0 \end{pmatrix}.$$

The finite $I_{xx}$ intensity computed for $N$TL ($N \geq 2$) (Table 1) is consistent with these Raman tensors. In our experiments, the signal is relatively weak for the mode near 176 cm$^{-1}$, but still can be observed for 2-5TL, under high energy excitation light such as 488 nm (Fig. 3a). For 6TL and above, this feature becomes too weak and merged into the background noise (Fig. 3a). Our calculations show that when the thickness reaches 6TL, the mode near 176 cm$^{-1}$ evolves into 6 modes with 3 of them Raman active. The 3 active Raman modes have a frequency range of 0.7 cm$^{-1}$, and the intensity of each sub-peak is only about 20% of that in 2TL WSe$_2$, thus explaining the difficulty in observing these peaks in the experiment. We further note that these Raman peaks near 176 cm$^{-1}$ are not related to substrate effects, as clearly shown by our spectra for suspended samples (Fig. 3b). The same conclusion applies to the peak at 310 cm$^{-1}$ discussed above (Fig. 3d).

We finally discuss the frequency trends as a function of thickness, focusing on the $A_{1g}$, $B_{2g}^1$ and $E_{2g}^1$ modes. Our first principles calculations predict that going from the bulk to 1TL, the $A_{1g}$ mode red-shifts by about 1 cm$^{-1}$, while the $B_{2g}^1$ and $E_{2g}^1$ modes blue-shift by about 3 cm$^{-1}$ and 2 cm$^{-1}$ respectively. The predicted blue-shift of the $B_{2g}^1$ mode (~2 cm$^{-1}$ from bulk to 2TL) agrees very well to the experimental measurements (Fig. 2b). While the $A_{1g}$ and $E_{2g}^1$ peaks cannot be resolved experimentally, we note that the $A_{1g}$ frequency is significantly higher than the $E_{2g}^1$ frequency in other TMD materials, such as MoS$_2$,[7,20,22,34-36] MoSe$_2$[7,20,36] and WS$_2$,[36] where the same predicted frequency trends have been observed. The relevant inter-atomic force constants for WSe$_2$ are very similar to those for MoS$_2$, and as such, the fact that $A_{1g}$ and $E_{2g}^1$ frequencies are so close in WSe$_2$ is related to the masses of W and Se.

In the nearest neighbor interaction picture, one would expect that a thinner film will result in less restoring force if the neighboring layers are vibrating in opposite directions. Based on the above intuitive understanding, it is expected that the frequencies of the modes here, in which the atoms in adjacent TLs are moving out-of-phase (180°), would red-shift with decreasing thickness (we call this the "thickness effect"). In previous work,[9] we used both first principles calculations and derived force-constants models to attribute the anomalous blue-shift of the $E_{2g}^1$ mode to the dominance of increased surface force constants over the thickness effect. Here, we show that the blue-shifts of both the $B_{2g}^1$ and $E_{2g}^1$ modes also result from this dominant surface effect. Specifically, following the procedure in Ref. 9, we construct force constants models (FCM) based on the LDA-calculated frequencies. Just as for $MoS_2$, we find that the force constants at the surface of the thin films differ significantly from those in the interior of the thin films and in the bulk (we call this the "surface effect"), with the W-Se force constants being significantly larger at the surface. To distinguish the thickness effect from the surface effect, we construct two models: Model 1 uses the force constants from bulk to construct the dynamical matrices; Model 2 is the same as Model 1 but uses modified surface force constants on both surfaces to simulate the surface effect. The results are shown in Figure 4b: without the surface effect, all the modes show a red-shift in frequencies with decreasing thickness, consistent with the intuitive understanding described above, while the surface effect leads to a blue-shift of the frequencies in $N$TL $WSe_2$. The thickness effect is most significant in the $A_{1g}$ mode, while the surface effect is most significant in the $B_{2g}^1$ and $E_{2g}^1$ mode. The competition between these two effects results in the red-shift of the $A_{1g}$ frequencies but the blue-shift of the $B_{2g}^1$ and $E_{2g}^1$ frequencies. We note that in the $E_{2g}^1$ mode, the surface displacements are negligible for $N$TL ($N > 2$) (Fig. 4b), thus explaining why the corresponding plot for Model 2 shows a sharp jump going from 3TL to 2TL (Fig. 4b). On the other hand, the surface displacements are relatively large for all thicknesses of the thin films in the case of the $B_{2g}^1$ mode (Fig. 4c). Therefore, there is a significant frequency difference of the $B_{2g}^1$ mode calculated by Models 1 and 2 for all the $N$TL WSe$_2$ (Fig. 4b). There is also a discontinuity between the frequencies for $N$TL and for the bulk in both the LDA calculations (Fig. 4a) and Model 2 (Fig. 4b). This discontinuity is directly related to the surface effect; no discontinuity is present in Model 1 where the surface effect is excluded.

## IV. CONCLUSIONS

In conclusion, we have found that reduced symmetry and dimensionality in 2D layered TMDs result in new Raman peaks that do not correspond to Raman-active peaks in the bulk, such as the new peaks around 176 cm$^{-1}$ and 310 cm$^{-1}$ under the $\bar{z}(xx)z$ polarized configuration in $N$TL ($N \geq 2$) WSe$_2$. These peaks correspond to the $E_{1g}$ and $B_{2g}^1$ modes in the bulk which have zero Raman tensor elements in our experimental configuration. Two major peaks near 250 cm$^{-1}$ and 260 cm$^{-1}$ are observed in our Raman scattering experiments. We have clarified that the peak near 250 cm$^{-1}$ is due to both $E_{2g}^1$ and $A_{1g}$ modes while the peak at 260 cm$^{-1}$ cannot be an out-of-plane vibrational mode, but may come from second-order Raman scattering processes. We further discuss the origin of the anomalous blue shift of the $E_{2g}^1$ and $B_{2g}^1$ modes with decreasing thickness, and found that the competition between the thickness effect and the surface effect plays an important role in the evolution behavior of these Raman frequencies. The findings we report here are likely to exist also in other 2D layered compounds with van der Waals interaction, such as other TMD materials, $Bi_2Te_3$ and $Bi_2Se_3$.

**Acknowledgements**

S.Y.Q. gratefully acknowledges support from the Institute of High Performance Computing Independent Investigatorship and from the Singapore National Research Foundation through a fellowship grant (NRF-NRFF2013-07). Q.X. gratefully acknowledges support from the Singapore National Research Foundation through a fellowship grant (NRF-RF2009-06). This work was also supported in part by Ministry of Education via a Tier 2 grant (MOE2012-T2-2-086) and start-up grant support (M58113004) from Nanyang Technological University (NTU). The authors acknowledge the support from the Singapore A*STAR Computational Resource Center.

Table 1. LDA calculated Γ point phonon frequencies (cm$^{-1}$) and relative Raman intensities $I_{xx}$ (shown in parentheses) of the phonon modes in 1-4TL WSe$_2$. The intensity is normalized by the largest value in each row. The irreducible representation and Raman [R] /Infrared [I] activity are also indicated; I+R indicates the mode is both Raman and Infrared active while Ina indicates that the mode is optically inactive. The bold font highlights the experimentally observed new modes that have non-zero $I_{xx}$ intensity in thin films but zero $I_{xx}$ intensity in the bulk. Frequencies for 5-7TL WSe$_2$ are shown in Fig. 4a.

| | | | | | | | | | | | | |
|---|---|---|---|---|---|---|---|---|---|---|---|---|
| In-plane Modes | bulk | 0 (0) $E^1_{1u}$ [I] | 23.76 (0.47) $E^2_{2g}$ [R] | 174.67 (0) $E_{2u}$ [Ina] | 176.31 (0) $E_{1g}$ [R] | 247.93 (1.0) $E^1_{2g}$ [R] | 248.17 (0) $E^2_{1u}$ [I] | | | | | |
| | 1TL | 0 (0) $E'$ [I+R] | 175.93 (0) $E''$ [R] | 250.24 (1.0) $E'$ [I+R] | | | | | | | | |
| | 2 TL | 0 (0) $E_u$ [I] | 17.08 (0.4) $E_g$ [R] | 175.42 (0) $E_u$ [I] | **176.26** **(0.012)** $E_g$ **[R]** | 249.19 (1.0) $E_g$ [R] | 249.31 (0) $E_u$ [I] | | | | | |
| | 3 TL | 0 (0) $E'$ [I+R] | 11.92 (0) $E''$ [R] | 21.28 (0.46) $E'$ [I+R] | 175.17 (0) $E''$ [R] | **175.84** **(0.016)** $E'$ **[I+R]** | 176.38 (0) $E''$ [R] | 248.21 (1.0) $E'$ [I+R] | 249.18 (0) $E''$ [R] | 249.20 (0.91) $E'$[I+R] | | | |
| | 4 TL | 0 (0) $E_u$ [I] | 9.51 (0.02) $E_g$ [R] | 17.26 (0) $E_u$ [I] | 22.72 (0.64) $E_g$ [R] | 175.03 (0) $E_u$ [I] | **175.52** **(0.010)** $E_g$ **[R]** | 176.06 (0) $E_u$ [I] | 176.40 (0.004) $E_g$ [R] | 248.11 (1.0) $E_g$ [R] | 248.23 (0) $E_u$ [I] | 249.20 (0.75) $E_g$ [R] | 249.19 (0) $E_u$ [I] |
| Out-of-plane Modes | bulk | 0 (0) $A^1_{2u}$ [I] | 37.60 (0.0) $B^2_{2g}$ [Ina] | 249.58 (0) $B_{1u}$ [Ina] | 252.27 (1.0, 0) $A_{1g}$ [R] | 307.66 (0) $A^2_{2u}$ [I] | 309.91 (0) $B^1_{2g}$ [Ina] | | | | | |
| | 1TL | 0 (0) $A''_2$ [I] | 251.32 (1.0) $A'_1$ [R] | 312.55 (0) $A''_2$ [I] | | | | | | | | |
| | 2 TL | 0 (0) $A_{2u}$ [I] | 27.01 (0.06) $A_{1g}$ [R] | 250.48 (0) $A_{2u}$ [I] | 251.77 (1.0) $A_{1g}$ [R] | 310.75 (0) $A_{2u}$ [I] | **311.26** **(0.0001)** $A_{1g}$ **[R]** | | | | | |
| | 3 TL | 0 (0) $A''_2$ [I] | 19.38 (0.07) $A'_1$ [R] | 33.49 (0) $A''_2$ [I] | 250.17 (0.02) $A'_1$ [R] | 251.17 (0) $A''_2$ [I] | 252.13 (1.0) $A'_1$ [R] | 309.54 (0) $A''_2$ [I] | **310.99** **(0.0001)** $A'_1$ **[R]** | 310.96 (0) $A''_2$ [I] | | | |
| | 4 TL | 0 (0) $A_{2u}$ [I] | 14.56 (0.07) $A_{1g}$ [R] | 27.11 (0) $A_{2u}$ [I] | 35.58 (0.003) $A_{1g}$ [R] | 250.00 (0) $A_{2u}$ [I] | 250.72 (0.06) $A_{1g}$ [R] | 251.55 (0) $A_{2u}$ [I] | 252.22 (1.0) $A_{1g}$ [R] | 309.26 (0) $A_{2u}$ [I] | 309.87 (0.00001) $A_{1g}$ [R] | 310.91 (0) $A_{2u}$ [I] | **310.98** **(0.0001)** $A_{1g}$ **[R]** |

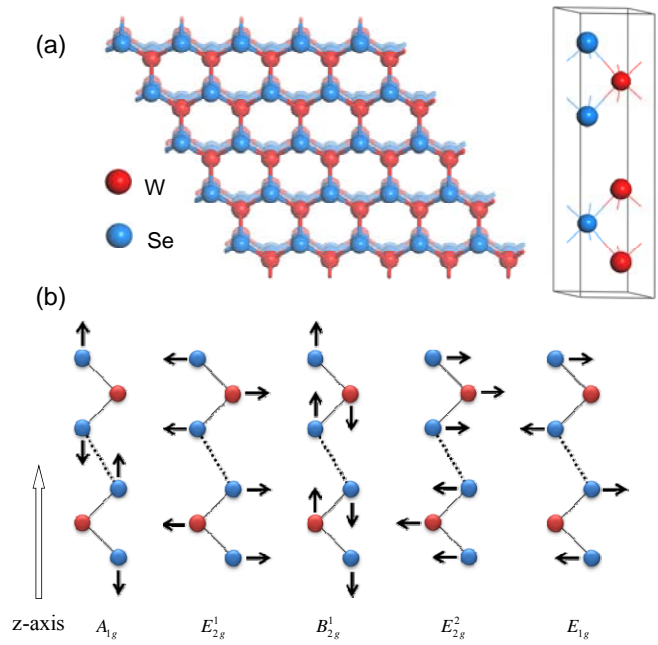

Figure 1. (a) Left: Tilted top view of 2TL WSe$_2$; Right: Side view of primitive cell of bulk 2H-WSe$_2$, (b) Side view of atomic displacements for the $A_{1g}$, $E_{2g}^1$, $B_{2g}^1$, $E_{2g}^2$ and $E_{1g}$ vibration modes.

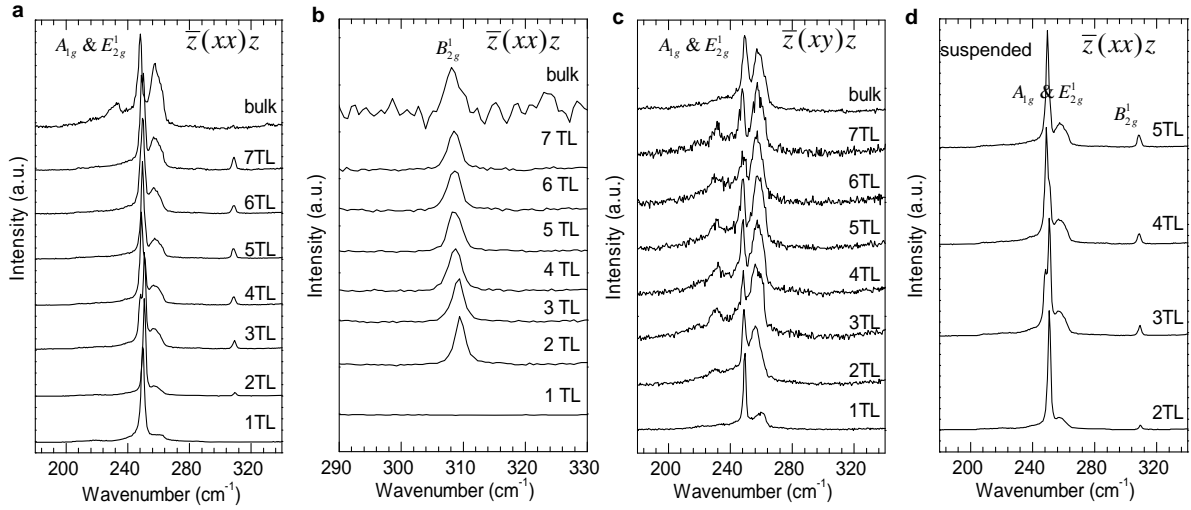

Figure 2. Raman spectra of 1-7TL WSe$_2$ under 532 nm excitation (a) under $\bar{z}(xx)z$ polarization configuration and (c) under $\bar{z}(xy)z$ polarization configuration. (b) is the zoom-in of (a) around 310 cm$^{-1}$. (d) Raman spectra for suspended 2-5TL WSe$_2$ under $\bar{z}(xx)z$ polarization configuration.

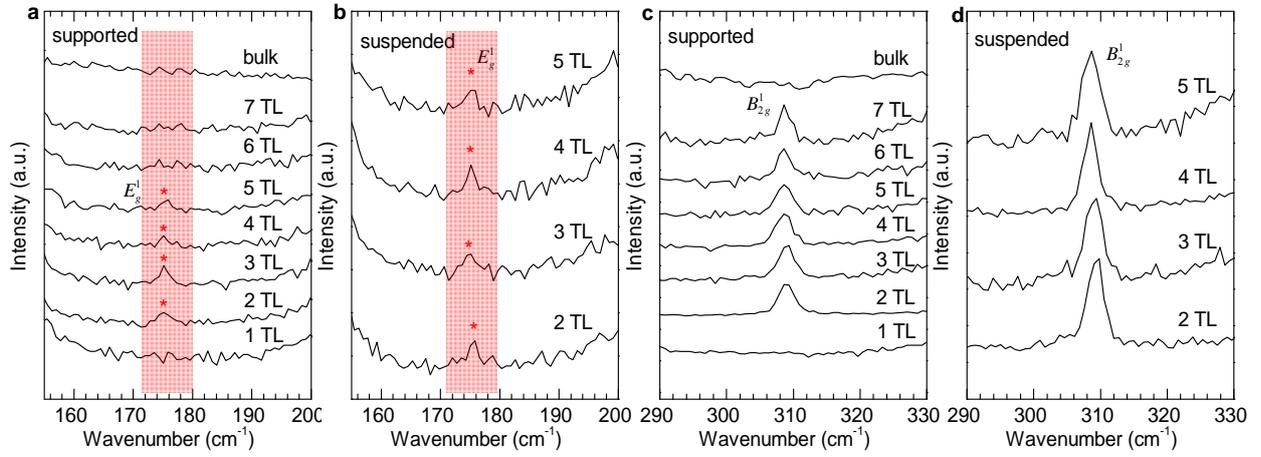

Figure 3. Raman spectra of 1-7TL and bulk WSe$_2$ under 488 nm excitation. (a) demonstration of the Raman peak located around 176 cm$^{-1}$ in 2-5TL WSe$_2$ supported by the SiO$_2$/Si substrates. (b) The same peak as in (a) from suspended samples. The Raman peak located around 310 cm$^{-1}$ is demonstrated in supported (c) and suspended (d) few-TL WSe$_2$.

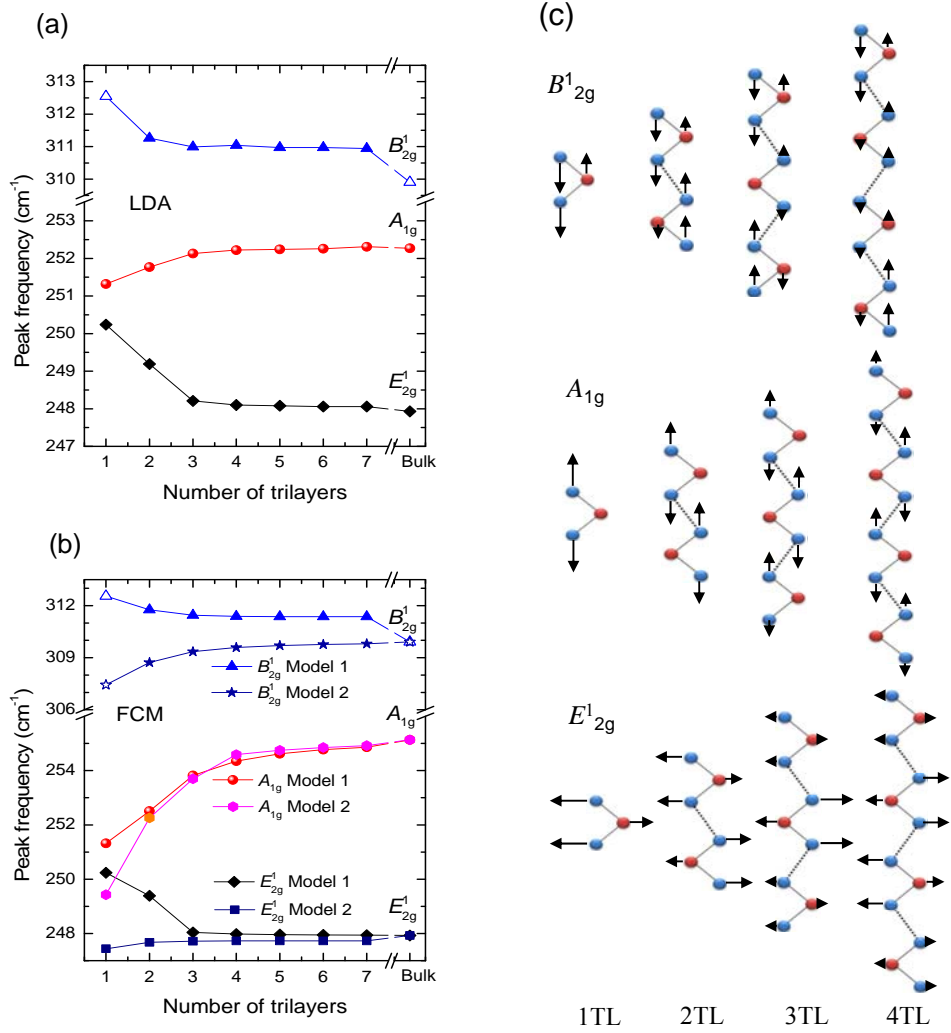

Figure 4. (a) DFT-LDA calculated phonon frequencies in bulk and 1-7TL WSe$_2$. The $B_{2g}^1$ mode in bulk and 1TL is not Raman active, and are plotted with empty symbols. (b) The force constants model (FCM) predicted phonon frequency evolution as a function of thickness. Model 1 uses parameters from the bulk, and Model 2 is the same as Model 1 but taking into account the surface effect by modifying surface force constants on both surfaces. (c) Schematic of the atomic displacements in $B_{2g}^1$, $A_{1g}$ and $E_{2g}^1$ modes in 1-4TL.

Supplementary Information for

# Effects of Lower Symmetry and Dimensionality on Raman Spectra in 2D WSe$_2$

Xin Luo[1,†], Yanyuan Zhao[2,†], Jun Zhang[2], Minglin Toh[4], Christian Kloc[4], Qihua Xiong[2,3,*], Su Ying Quek[1,+,*]

[1]Institute of High Performance Computing, 1 Fusionopolis Way, #16-16 Connexis, Singapore 138632

[2]Division of Physics and Applied Physics, School of Physical and Mathematical Sciences, Nanyang Technological University, 21 Nanyang Link, Singapore 637371

[3]Division of Microelectronics, School of Electrical and Electronic Engineering, Nanyang Technological University, Singapore, 639798

[4]School of Materials Science and Engineering, Nanyang Technological University, Nanyang Avenue, Singapore 639798

*: To whom correspondence should be addressed. Email address: phyqsy@nus.edu.sg (theory); qihua@ntu.edu.sg (experiment)
+: Present address: Department of Physics, Graphene Research Centre and Centre for Computational Science and Engineering, National University of Singapore, 2 Science Drive 3, Singapore 117551
†: These authors contributed equally to this work.

**Contents**
-$E^1_{2g}$ **and** $A_{1g}$ **modes under 633 nm excitation**
-$B^1_{2g}$ **mode evolution from 1Trilayer to multi Trilayer**

# $E^1_{2g}$ and $A_{1g}$ modes under 633 nm excitation

The spectral resolution of our spectrometer is 1 cm$^{-1}$, which should not be a problem for distinguishing the $E_{2g}^1$ and $A_{1g}$ modes. However, experimentally, these two modes cannot be distinguished in thin films under the 532 nm or 488 nm excitations mainly because that the Raman intensity of the $E_{2g}^1$ mode is much weaker than that of the $A_{1g}$ mode. In the bulk crystal, the intensity difference is smaller and thus the two peaks can be distinguished. Under 633 nm excitation, the $E_{2g}^1$ and $A_{1g}$ modes can be distinguished in few layers because of a smaller intensity difference (see Figure S1). The frequency differences between these two modes are 2.7 cm$^{-1}$ for bulk and decreases to 1.1 cm$^{-1}$ in 2TL. In 1TL, the difference is even smaller and thus the two peaks cannot be experimentally distinguished. Similar results were also reported by W. Zhao *et al.* (arXiv:1304.0911).

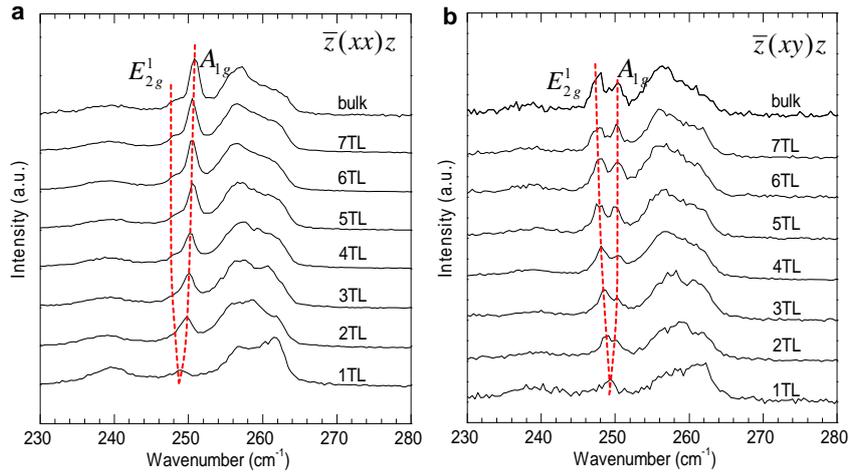

Figure S1. Raman spectra of 1-7TL and bulk WSe$_2$ under 633 nm excitation under (a) $\bar{z}(xx)z$ polarization configuration and (b) $\bar{z}(xy)z$ polarization configuration. The red dashed lines indicate the blue shift and red shift trends of the $E^1_{2g}$ and $A_{1g}$ modes, respectively, with decreasing thickness.

Note that the finite intensity of the $A_{1g}$ mode under $\bar{z}(xy)z$ polarization configuration (Figure S1b) is due to the experimental limit. Experimentally, we use an analyzer (polarizer) to separately detect the scattered light with xx and xy polarizations. Weather we can see the $A_{1g}$ mode under $\bar{z}(xy)z$ configuration is determined by how efficient the analyzer is. In our setup, the efficiency of the analyzer is 90%, that is to say, 10% of the $A_{1g}$ mode intensity can still be detected even under $\bar{z}(xy)z$ configuration, which explains the finite intensity of the $A_{1g}$ mode under $\bar{z}(xy)z$ polarization. Because of the experimental limit, we distinguish the polarization of a Raman mode through its relative intensity compared to other modes, instead of the absolute intensity. For example, relative to the $E^1_{2g}$ mode, the $A_{1g}$ mode distinctively gets weakened under the $\bar{z}(xy)z$ polarization and we therefore assign them to in-plane and out-of-plane vibrational modes, respectively.

In our Raman measurements, we used a laser power of 0.3 mW for both 532 nm and 488 nm excitations. The integration time is 60 s for few layers and 200 s for the bulk crystal.

### $B^1_{2g}$ mode evolution from 1Trilayer (TL) to multi-Trilayer

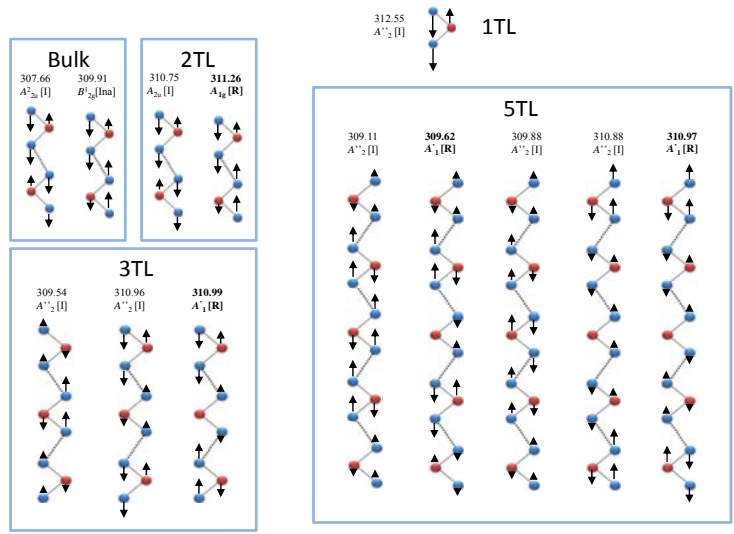

Figure S2. The $B^1_{2g}$ mode in bulk comes from the $A''_2$ mode in 1TL, and the same mode in 1TL evolves into N modes with similar frequencies in N TLs. Some of these modes are Raman active in N TLs (N>1), as shown in the bold.